\documentclass[10pt,reqno,twoside,paper=letter]{amsart}

\usepackage[letterpaper,left=20mm,right=14mm,layoutvoffset=10mm]{geometry}
\usepackage{epsfig,graphicx}

\usepackage[bf,small]{titlesec}
\titlelabel{\thetitle.\,\,\,}
\usepackage{amssymb,amsfonts}
\usepackage{multicol}
\usepackage{fancyhdr}
\usepackage[latin1]{inputenc}				
\usepackage[comma,longnamesfirst,sectionbib]{natbib}
\usepackage{enumerate}
\usepackage[symbol*]{footmisc}
\newcommand{\inforevista}{\scriptsize  Rev. Acad. Colomb. Cienc. Ex. Fis. Nat. nn(nnn):ww--zzz,ddd-ddd de 2016}
\begin{document}
%
%
\pagenumbering{arabic}
\fancypagestyle{plain}{%
\fancyhf{} 
\fancyfoot[R]{\thepage} %
\fancyhead[L]{\inforevista}
\renewcommand{\headrulewidth}{0pt}
\renewcommand{\footrulewidth}{0pt}}
\thispagestyle{plain} 
\pagestyle{fancy}     
\fancyhead{} 
\renewcommand{\headrulewidth}{0.0pt} 
\fancyhead[LO]{\inforevista}
\fancyhead[RO]{\scriptsize Distribution functions for a family of galaxy models}
\fancyhead[LE]{\scriptsize G. A. González,  J. F. Pedraza and J. Ramos-Caro}
\fancyhead[RE]{\inforevista}
\fancyfoot{} 
\fancyfoot[LE]{\thepage}
\fancyfoot[RO]{\thepage}

\begin{flushright}
\rule[-0.5ex]{0.5ex}{3.0ex} {\large Physical Sciences}
\end{flushright}
\vspace*{0.3cm}

\begin{center}
{\LARGE \textbf{Distribution functions for a family of axially symmetric galaxy models \\}}
\end{center}

\vspace{3mm}
\begin{center}
\textbf{\small Guillermo A. González${}^{1,}\footnotemark[1],$ Juan F. Pedraza${}^{2},$ and Javier Ramos-Caro${}^{3}$}
\vspace{3mm}

{\scriptsize
${}^{1}$Escuela de Física, Universidad Industrial
de Santander, Bucaramanga, Colombia\\
${}^{2}$Institute for Theoretical Physics, University of Amsterdam, Amsterdam, Netherlands\\
${}^{3}$Departamento de Física, Universidade Federal de S\~ao Carlos,
S\~ao Carlos, Brazil\\ }
\end{center}
\footnotetext[1]{Correspondence: G. A. González, guillermo.gonzalez@saber.uis.edu.co, Received xxxxx XXXX; Accepted xxxxx XXXX.}

\begin{Small}
\vspace{3.0mm}
\rule{\textwidth}{0.4pt}

\begin{center}
\begin{minipage}{14cm}
\vspace{3mm}

\textbf{Abstract}
\vspace{3mm}

{We present the derivation of distribution functions for the first four members of a family of disks, previously obtained in \cite{GR}, which represent a family of axially symmetric galaxy models with finite radius and well-behaved surface mass density. In order to do this, we employ several approaches that have been developed starting from the potential-density pair and, essentially using the method introduced by \cite{KAL2}, we obtain some distribution functions that depend on the Jacobi integral. Now, as this method demands that the mass density can be properly expressed as a function of the gravitational potential, we can do this only for the first four disks of the family. We also find another kind of distribution functions by starting with the even part of the previous distribution functions and using the maximum entropy principle in order to find the odd part and so a new distribution function, as it was pointed out by \cite{DEJ}. The result is a wide variety of equilibrium states corresponding to several self-consistent finite flat galaxy models.
\\[1mm]

}

\vspace{3mm}

\textbf{Funciones de distribución para una familia de modelos de galaxias axialmente simétricas}
\vspace{3mm}

\textbf{Resumen}
\vspace{3mm}

{Se presenta la derivación de funciones de distribución para los primeros cuatro miembros de una familia de discos, obtenida previamente en \cite{GR}, la cual representa a una familia de modelos de galaxias axialmente simétricas de radio finito y con densidad superficial de masa bien comportada. Para ello, se emplean varios enfoques desarrollados a partir del par potencial-densidad y, utilizando esencialmente el método introducido por \cite{KAL2}, se obtienen algunas funciones de distribución que dependen de la integral de Jacobi. Ahora, ya que este método exige que la densidad de masa se pueda expresar adecuadamente como una función del potencial gravitacional, sólo es posible hacer esto para los primeros cuatro discos de la familia. También encontramos otro tipo de funciones de distribución, comenzando con la parte par de las funciones de distribución anteriores y utilizando el principio de máxima entropía con el fin de encontrar la parte impar y por lo tanto una nueva función de distribución, como fue señalado por \cite{DEJ}. El resultado es una amplia variedad de estados de equilibrio correspondiente a varios modelos auto-consistentes de galaxias planas finitas.
\\[1mm]

\textbf{Palabras clave:} Dinámica estelar, Galaxias: cinemática y dinámica.}
\end{minipage}
\end{center}

\rule{\textwidth}{0.4pt}
\end{Small}

\begin{small}
\columnsep 0.5 cm
\begin{multicols}{2}

\setlength{\parskip}{.3cm}

\section*{Introduction}

The problem of finding self-consistent stellar models for galaxies is of wide interest in astrophysics. Usually, once the potential-density pair (PDP) is formulated as a model for a galaxy, the next step is to find the corresponding distribution function (DF). This is one of the fundamental quantities in galactic dynamics specifying the distribution of the stars in the phase-space of positions and velocities. Although the DF can generally not be measured directly, there are some observationally accesible quantities that are closed related to the DF: the projected density and the line-of-sight velocity, provided by photometric and kinematic observations, are examples of DF moments. Thus, the formulation of a PDP with its corresponding equilibrium DFs establish a self-consistent stellar model that can be corroborated by astronomical observations.

On the other hand, a fact that is usually assumed in astrophysics, see \cite{BT}, is that the main part of the mass of a typical spiral galaxy is concentrated in a thin disk. Accordingly, the study of the gravitational potential generated by an idealized thin disk is a problem of great astrophysical relevance and so, through the years, different approaches has been used to obtain the PDP for such kind of thin disk models (see \cite{KUZ} and \cite{T1,T2}, as examples).

Now, a simple method to obtain the PDP of thin disks of finite radius was developed by \cite{HUNT}, the simplest example of disk obtained by this method being the \cite{KAL1} disk. In \cite{GR}, we use the Hunter method in order to obtain an infinite family of axially symmetric finite thin disks, characterized by a well-behaved surface density, whose first member is precisely the well-known Kalnajs disk. Also, the motion of test particles in the gravitational fields generated by the first four members of this family was studied in \cite{RLG}, and a new infinite family of self-consistent models was obtained in \cite{PRG} as a superposition of members belonging to the family.

We will consider at the present paper the derivation of some two-integral DFs for the first four members of the family obtained in \cite{GR}. Now, as is stated by the Jeans theorem, an equilibrium DF is a function of the isolating integrals of motion that are conserved in each orbit and, as it has been shown, it is possible to find such kind of DFs for PDPs such that there is a certain relationship between the mass density and the gravitational potential. The simplest case of physical interest corresponds to spherically symmetric PDPs, which are  described by isotropic DFs that depends on the total energy $E$. Indeed, as was be shown in \cite{EDD}, it is possible to obtain this kind of isotropic  DFs by first expressing the density as a function of the potential and then solving an Abel integral equation.

\cite{JO} found that a similar procedure can be performed in the axially symmetric case, where the equilibrium DF depends on the energy $E$ and the angular momentum about the axis of symmetry $L_{z}$, i.e. the two classical integrals of motion. They developed a formalism that essentially combines both the Eddington formulae and the \cite{FRICKE} expansion in order to obtain the DF even part, starting from a density that can be expressed as a function of the radial coordinate and the gravitational potential. Once such even part is determined, the DF odd part can be obtained by introducing some reasonable assumptions about the mean circular velocity or using the maximum entropy principle.

On the other hand, another method appropriated to find DFs depending only of the Jacobi integral, $E_r = E - \Omega L_z$, for axially symmetric flat galaxy models was introduced by \cite{KAL2} for the case of disk-like systems and basically consists in to express the DF as a derivative of the surface mass density with respect to the gravitational potential. Such method does not demands solving an integral equation, but instead is necessary properly to express the mass density as a function of the gravitational potential, a procedure that is only possible in some cases.

In this paper we will use both of the mentioned approaches in order to obtain equilibrium DFs for some generalized Kalnajs disks. Accordingly, the paper is organized as follows. First, we present the fundamental aspects of the two methods that we will use in order to obtain the equilibrium DFs. Then, we present a summary of the main aspects of the generalized Kalnajs disks, and then we derive the DFs for the first four members of the family. Finally, we summarize our main results.

\section*{Formulation of the Methods}\label{sec:finteq}

We assume that $\Phi$ and $E$ are, respectively, the gravitational potential and the energy of a star in a stellar system. One can choose a constant $\Phi_0$ such that the system has only stars of  energy $E < \Phi_0,$ and then define a relative potential $\Psi = - \Phi + \Phi_0$ and a relative energy $\varepsilon= - E + \Phi_0$, see \cite{BT}, such that $\varepsilon=0$ is the energy of escape from the system. Both the mass density $\rho({\bf r})$ and the DF $f({\bf r},{\bf v})$ are related to $\Psi({\bf r})$ through the Poisson equation
\begin{equation}
\nabla^2\Psi = - 4\pi G \rho = -4\pi G\int fd^3{\bf v}, \label{int0}
\end{equation}
where $G$ is the gravitational constant.

For the case of an axially symmetric system, it is customary to use cylindrical polar coordinates $(R,\varphi,z)$, where ${\bf v}$ is denoted by ${\bf v}=(v_R,v_\varphi,v_z)$. As it is well known, such system admits two isolating integrals for any orbit: the component of the angular momentum about the$z$-axis, $L_z=R v_\varphi$, and the relative energy $\varepsilon$. Hence, by the Jeans theorem, the DF of a steady-state stellar system in an axially symmetric potential can be expressed as a non-negative function of $\varepsilon$ and $L_z$, denoted by $f(\varepsilon,L_z)$. Such DF, that vanishes for $\varepsilon<0$, is related to the mass density through eq. (\ref{int0}).

In this subject, $f(\varepsilon,L_z)$ is usually separated into even and odd parts, $f_{+}(\varepsilon,L_z)$ and $f_{-}(\varepsilon,L_z)$ respectively, with respect to the angular momentum $L_z$, where
\begin{equation}
f_+(\varepsilon,L_z) = \begin{matrix} \frac{1}{2}\end{matrix}
[f(\varepsilon,L_z) + f(\varepsilon,-L_z)],
\end{equation}
and
\begin{equation}
f_-(\varepsilon,L_z) = \begin{matrix} \frac{1}{2}\end{matrix}
[f(\varepsilon,L_z) - f(\varepsilon,-L_z)].
\end{equation}
So, by using 
\begin{equation}
\varepsilon=\Psi-\begin{matrix}
\frac{1}{2}\end{matrix}(v_R^2+v_\varphi^2+v_z^2),
\end{equation}
the integral given by (\ref{int0}) can be expressed, see \cite{BT}, as
\begin{equation}
\rho = \frac{4\pi}{R} \int_0^\Psi \int_0^{R \sqrt{2(\Psi - \varepsilon)}}
f_+(\varepsilon,L_z) dL_z d\varepsilon. \label{inta2}
\end{equation}
For a given mass density, this relation can be considered as the integral equation determining $f_+(\varepsilon,L_z)$, while the odd part satisfies the relation
\begin{equation}
 \rho R \langle v_{\varphi}\rangle =
\frac{4\pi}{R}\int_0^\Psi \int_0^{R\sqrt{2(\Psi-\varepsilon)}}
L_{z}f_-(\varepsilon,L_z)dL_z d\varepsilon. \label{DFgen2}
\end{equation}
This integral equation was first found by \cite{LB} and then applied by \cite{EVANS} into calculating the odd DF for the Binney model, under the assumption of $\langle v_{\varphi}\rangle$ having some realistic rotational laws.

As $\langle v_{\varphi}\rangle$ is not known, we cannot compute
$f_-(\varepsilon,L_z)$ directly by eq. (\ref{DFgen2}) but what we can do is to
obtain the most probable distribution functions under some suitable assumptions.
Once $f_+(\varepsilon,L_z)$ is known, $f_{-}(\varepsilon,L_z)$, and therefore
$f(\varepsilon,L_z)$, can be obtained by means of the maximum entropy principle, see
\cite{DEJ}, and we obtain
\begin{equation}
f_{-}(\varepsilon,L_z)=f_+(\varepsilon,L_z)\frac{e^{\alpha L_{z}}-1}{e^{\alpha
L_{z}}+1},
\end{equation}
\begin{equation}
f(\varepsilon,L_z) = \frac{2f_+(\varepsilon,L_z)}{1 + e^{-\alpha L_{z}}},
\label{DFgen1}
\end{equation}
where $\alpha$ is the parameter depending on the total angular momentum.
Obviously, $|f_{-}(\varepsilon,L_z)|\leq f_{+}(\varepsilon,L_z)$. Also, the
system is non-rotating when $\alpha=0$ and maximally rotating as
$\alpha\rightarrow\infty$, i.e., for $\alpha\rightarrow +\infty$, it is
anticlockwise and
$f(\varepsilon,L_z)=[1 + \mathrm{sign}(L_z)]f_{+}(\varepsilon,L_z)$, for
$\alpha\rightarrow - \infty$, the rotation is clockwise and
$f(\varepsilon,L_z)=[1-\mathrm{sign}(L_z)]f_{+}(\varepsilon,L_z)$. The parameter
$\alpha$ reflects the rotational characteristics of the system.

As it was pointed out by \cite{FRICKE} and recently by \cite{JO}, the
implementation of integral equations (\ref{inta2}) and (\ref{DFgen2}) demands
that one can express $\rho$ as a function of $R$ and $\Psi$. This holds indeed
for the case of disk-like systems, which surface mass density $\Sigma$ is
related to $f$ through
\begin{equation}
\Sigma = 4 \int_0^\Psi \int_0^{R\sqrt{2(\Psi - \varepsilon)}}
\frac{f_+(\varepsilon,L_z)dL_z}{\sqrt{2R^{2}(\Psi - \varepsilon)-L_{z}^{2}}}
d\varepsilon. \label{inta3}
\end{equation}
In order to incorporate the formalisms developed for the 3-dimensional case to
deal with disk-like systems, we have to generate a pseudo-volume density
$\hat{\rho}$, see \cite{HQ}, according to
\begin{equation}
\hat{\rho}= \sqrt{2} \int_{0}^{\Psi} \frac{\Sigma(R^{2},\Psi')
 d\Psi'}{\sqrt{\Psi - \Psi'}}, \label{seudorho}
\end{equation}
which must take the place of $\rho$ in (\ref{inta2}) and (\ref{DFgen2}). In
particular, when $\hat{\rho}(\Psi,R) = \sum\limits_{n=0}^{m} R^{2n}
\varrho_n(\Psi)$, see  \cite{JO}, the corresponding even DF is
\begin{eqnarray}
f_+(\varepsilon,L_z) &=& \sum\limits_{n=0}^{m} \frac{2^{- n - 3/2}
L_z^{2n}}{\pi^{3/2} \Gamma(n + \frac{1}{2})} \nonumber \\
&&\times \left[ \int_0^\varepsilon
\frac{d^{n + 2} \varrho_n(\Psi)}{d\Psi^{n + 2}} \frac{d\Psi}{\sqrt{\varepsilon -
\Psi}} \right. \nonumber\\
&&\ \left. + \frac{1}{\sqrt{\varepsilon}}
\left(\frac{d^{n + 1} \varrho_n(\Psi)}{d\Psi^{n+1}} \right)_{\Psi=0} \right].
\label{JO}
\end{eqnarray}

Another simpler method to find equilibrium DFs corresponding to axially
symmetric disk-like systems, was introduced by \cite{KAL2}. Such formalism deal
with DFs that depend on the Jacobi's integral $E_r=E-\Omega L_z$, i.e. the
energy measured in a frame rotating with constant angular velocity $\Omega$. It
is convenient to define an effective potential $\Phi_{r}=\Phi-\begin{matrix}
\frac{1}{2}\end{matrix}\Omega^2R^{2}$ in such way that, if we choose a frame in
which the velocity distribution is isotropic, the DF will be $L_{z}$-independent
and, from (\ref{inta3}), the relation between the surface mass density and the
DF is reduced to
\begin{equation}
\Sigma=
2 \pi \int_{0}^{\Psi_{r}} f(\varepsilon_{r}) d\varepsilon_{r}. \label{metkal}
\end{equation}
Here, $\Psi_{r} = - \Phi + \begin{matrix}\frac{1}{2}\end{matrix} \Omega^2 R^{2}
+ \Phi_{0r}$ and $\varepsilon_{r} = \varepsilon + \Omega L_{z} + \Phi_{0r} -
\Phi_{0}$, i.e. the relative potential and the relative energy measured in the
rotating frame. Moreover, if one can express $\Sigma$ as a function of
$\Psi_{r}$, differentiating both sides of (\ref{metkal}) with respect to
$\Psi_{r}$, we obtain
\begin{equation}
f(\varepsilon_{r}) = \left. \frac{1}{2\pi} \frac{d\Sigma}{d\Psi_{r}}
\right|_{\Psi_{r} = \varepsilon_{r}}.  \label{metkal2}
\end{equation}
Note that in this formalism it is also necessary to express the mass density as
a function of the relative potential.

\section*{DFs for the family of disks}\label{sec:dfkal}

\subsection*{The family of disk models.}\label{sec:kal}

In \cite{GR}, we obtain an infinite family of axially symmetric
finite thin disks such that the mass surface density of
each model (labeled with the positive integer $m \geq 1$) is given by
\begin{equation}
\Sigma_{m}(R) = \Sigma_{c}^{(m)}\left[1 - \frac{R^{2}}{a^{2}} \right]^{m-1/2},
\label{densidad}
\end{equation}
with the constants $\Sigma_{c}^{(m)}$ given by
\begin{equation}
\Sigma_{c}^{(m)}=\frac{(2m+1)M}{2\pi a^{2}},
\end{equation}
where $M$ is the total mass and $a$ is the disk radius. Such mass distribution
generates an axially symmetric gravitational potential, that can be written as
\begin{equation}
\Phi_{m}(\xi,\eta) = - \sum_{n=0}^{m} C_{2n} q_{2n}(\xi) P_{2n}(\eta) \label{dk}
\end{equation}
where $P_{2n}(\eta)$ and $q_{2n}(\xi) = i^{2n+1} Q_{2n}(i\xi)$ are the usual
Legendre polynomials and the Legendre functions of the second kind respectively,
and $C_{2n}$ are constants given by
$$
C_{2n} = \frac{M G\pi^{1/2} (4n+1) (2m+1)!}{a2^{2m+1} (2n+1) (m - n)! \Gamma(m +
n + \frac{3}{2} )q_{2n+1}(0)},
$$
where $G$ is the gravitational constant. Here, $-1 \leq \eta \leq 1$ and $0 \leq
\xi < \infty$ are spheroidal oblate coordinates, related to the usual
cylindrical coordinates $(R,z)$ through the relations
\begin{equation}
R^{2} = a^{2}(1 + \xi^{2})(1 - \eta^{2}),\qquad z = a \eta \xi.\label{tco}
\end{equation}
In particular, we are interested in the gravitational potential at the disk,
where $z = 0$ and $0 \leq R \leq a$, so $\xi = 0$ and $\eta =
\sqrt{1-R^{2}/a^{2}}$. 

If we choose $\Phi_{0m}$ in such a way that $\Psi_{m} \geq 0$, the corresponding
relative potential for the first four members will be
\begin{subequations}\begin{align}
\Psi_{1}(R) &=  \frac{3\pi MG }{8a^{3}}(a^{2}-R^{2}),
\label{eq:4.22}   \\
\Psi_{2}(R) &= \frac{15\pi MG}{128a^{5}} (5a^{4}-8R^{2}a^{2}+3R^{4}), \label{eq:4.23}  \\
\Psi_{3}(R) &= \frac{35\pi MG}{512a^{7}} (11a^{6}-24R^{2}a^{4}\nonumber\\
&\qquad \qquad \qquad
\qquad+18R^{4}a^{2}-5R^{6}),\label{eq:4.24} \\
\Psi_{4}(R) &= \frac{315\pi MG}{32768a^{9}}
(93a^{8}-256R^{2}a^{6}+288R^{4}a^{4}\nonumber\\
&\qquad \qquad \qquad
\qquad-160R^{6}a^{2}+35R^{8}),\label{eq:4.25}
\end{align}\end{subequations}
We shall restrict our attention to these four members. The formulae
showed above defines the relative potentials that will be used to
calculate the DFs by the implementation of the methods sketched in
previous section.

\subsection*{DFs for the $m=1$ disk.}\label{sec:dfm1}

Given the mass surface distribution (\ref{densidad}) and the relative potential
(\ref{eq:4.22}), we can easily obtain the following relation:
\begin{equation}
\Sigma_1(\Psi_1) = \frac{\sqrt{2} \Sigma_c^{(1)}}{\Omega_0a} \Psi_1^{1/2},
\label{dens1}
\end{equation}
where $\Omega_0=[3\pi GM/(4a^3)]^{1/2}$. As we are dealing with disk-like
systems, it is necessary to compute the pseudo-volume density $\hat{\rho}$ by
eq. (\ref{seudorho}) in order to perform the integral (\ref{inta2}),
\begin{equation}
\hat{\rho}_1(\Psi_1) = \frac{\pi \Sigma_c^{(1)}}{\Omega_0a} \Psi_1
\label{pseudo1}.
\end{equation}
By using the right part of eq. (\ref{JO}) with $n=0$, which is equivalent to
calculate the Fricke component of (\ref{pseudo1}), we obtain the even part of a
DF which depends only on the relative energy
\begin{equation}
f_{1+}^{(A)}(\varepsilon) = \frac{\Sigma_c^{(1)}}{2 \pi \Omega_0a
\sqrt{2\varepsilon}} \label{even1a}.
\end{equation}
At this point, we may notice that this $f_{1+}(\varepsilon)$ corresponds to the
DF formulated by \cite{BT} when $\Omega = 0$. (Note that there is a
difference of constants as a result of a different definition of the relative
potential and relative energy.)

\begin{figure*}
\centering
\epsfig{width=5.6in,file=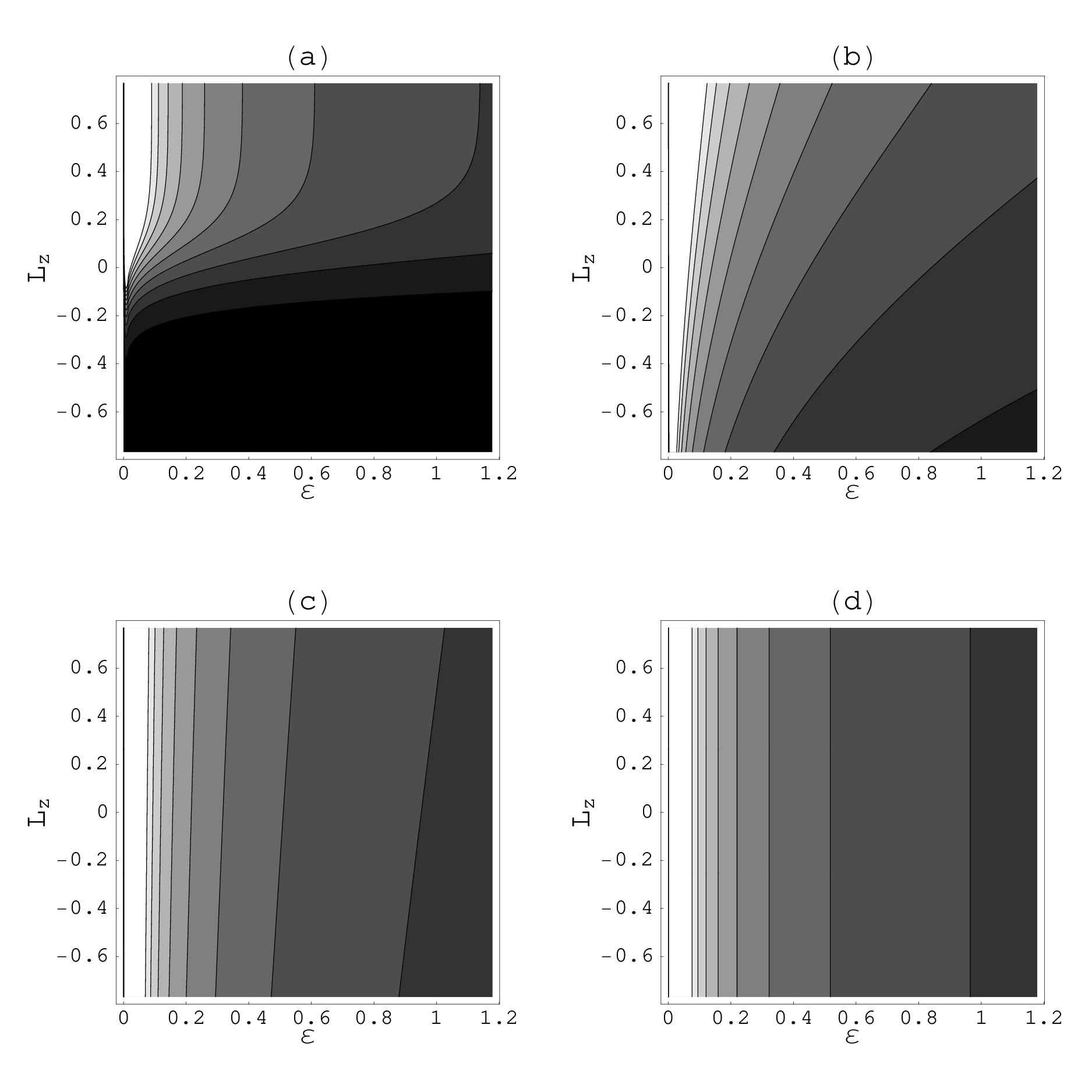}
\caption{Contours of $f_{1}^{(A)}$, given by (\ref{DFm1A}), for (a) $\alpha=10$,
(b) $\alpha=1$, (c) $\alpha=0.1$ and (d) $\alpha=0$. Larger values of the DF
corresponds to lighter zones.}\label{fig:DFm1A}
\end{figure*}

To obtain a full DF, we use the maximum entropy principle by means of eq.
(\ref{DFgen1}) and the result is
\begin{equation}
f_{1}^{(A)}(\varepsilon,L_{z}) = \frac{\Sigma_c^{(1)}}{\pi \Omega_0a
\sqrt{2\varepsilon}(1 + e^{-\alpha L_z})}.\label{DFm1A}
\end{equation}
In figure \ref{fig:DFm1A} we shown the contours of $f_{1}^{(A)}$ for different values of $\alpha$: we take $\alpha = 10$ in figure \ref{fig:DFm1A}(a), $\alpha = 1$ in figure \ref{fig:DFm1A}(b), $\alpha = 0.1$ in figure \ref{fig:DFm1A}(c) and $\alpha = 0$ in figure \ref{fig:DFm1A}(d). As it is shown in the figures, $\alpha$ determines a
particular rotational state in the stellar system (from here on we
set $G=a=M=1$ in order to generate the graphics, without loss of
generality). As $\alpha$ increases, the probability to find a star
with positive $L_{z}$ increases as well. A similar result can be
obtained for $\alpha<0$, when the probability to find a star with
negative $L_{z}$ decreases as $\alpha$ decreases, and the
corresponding plots would be analogous to figure \ref{fig:DFm1A},
after a reflection about $L_{z}=0$.

\begin{figure*}
\centering
\epsfig{width=5.6in,file=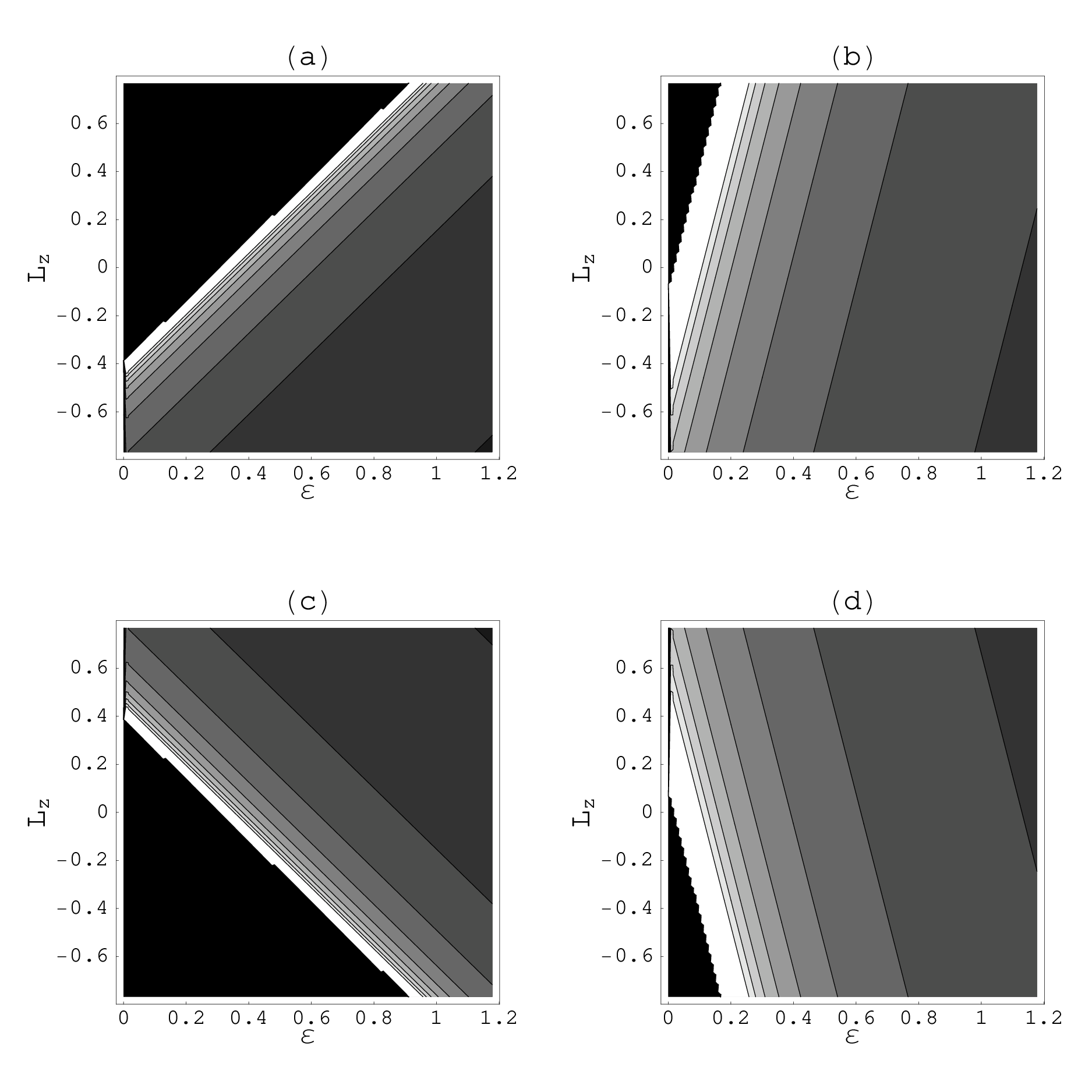}
\caption{Contours of $f_{1}^{(B)}$, given by (\ref{DFm1B1}), for (a)
$\Omega=-\pi/4$, (b) $\Omega=-\pi/16$, (c) $\Omega=\pi/4$ an (d) $\Omega=\pi/16$. Larger values of the DF corresponds to lighter zones.}\label{fig:DFm1C}
\end{figure*}

\begin{figure*}
\centering
\epsfig{width=5.6in,file=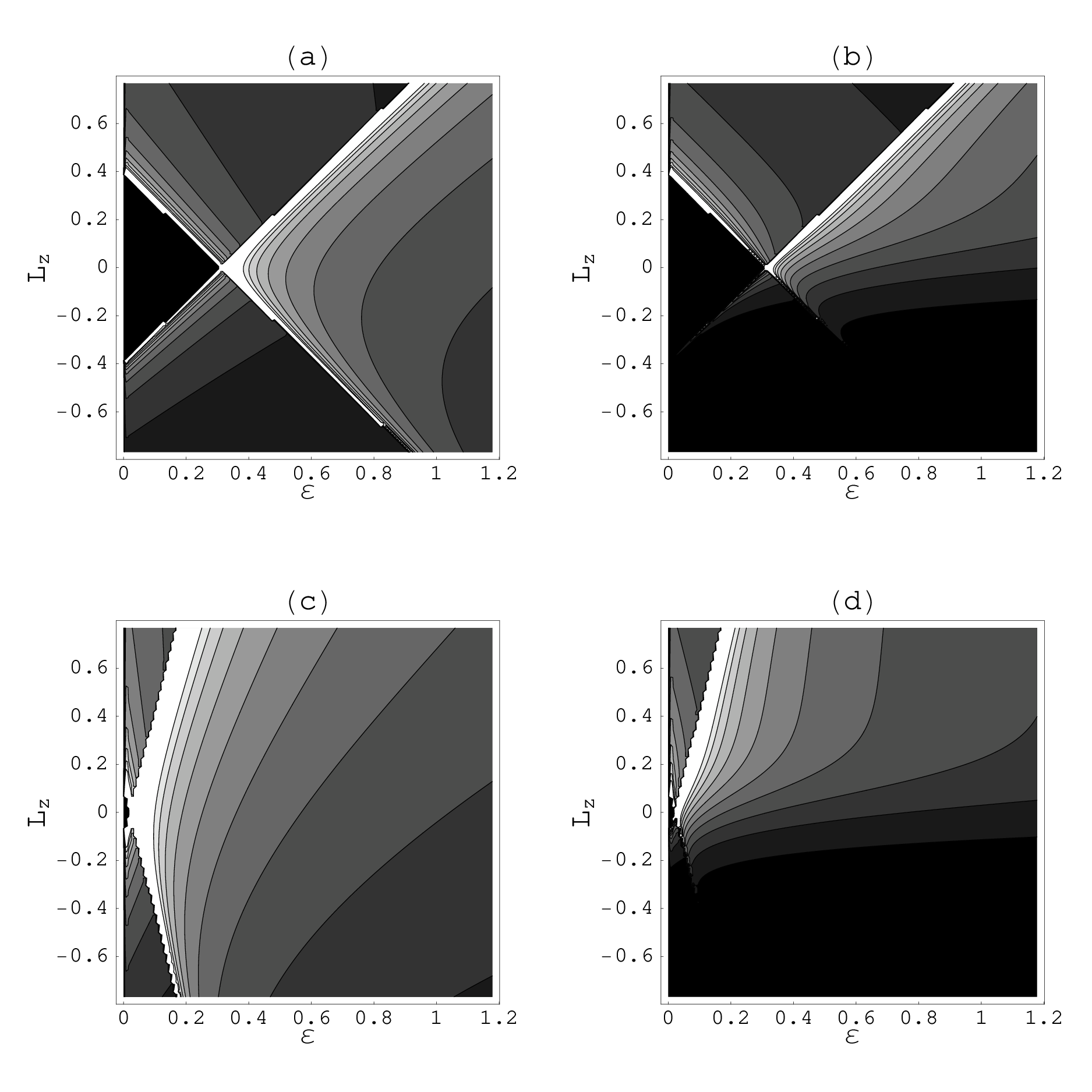}
\caption{Contours of $\tilde{f}_{1}^{(B)}$, given by (\ref{DFm1B}), for
$\Omega=\pi/4$ with (a) $\alpha=1$, and (b) $\alpha=10$. For $\Omega=\pi/16$ we have
(c) $\alpha=1$, and (d) $\alpha=10$. Larger values of the DF corresponds to lighter
zones.}\label{fig:DFm1B}
\end{figure*}

We can generalize this result if we perform the analysis in a rotating frame. At
first instance, it is necessary to deal with the effective potential in order to
take into account the fictitious forces. Choosing conveniently $\Phi_{0r}$, the
relative potential in the rotating frame takes the form
\begin{equation}
\Psi_{1r}(R) = \frac{(\Omega_0^2 - \Omega^2)}{2}(a^2 - R^2),
\end{equation}
so the corresponding mass surface density and the pseudo-volume density can be
expressed as
\begin{equation}
\Sigma_1(\Psi_{1r}) = \frac{\sqrt{2} \Sigma_c^{(1)}}{a \sqrt{\Omega_0^2 -
\Omega^2}} \Psi_{1r}^{1/2}
\end{equation}
and
\begin{equation}
\hat{\rho}(\Psi_{1r}) = \frac{\pi \Sigma_c^{(1)}}{a \sqrt{\Omega_0^2 -
\Omega^2}} \Psi_{1r}.
\end{equation}
The resulting even part of the DF in the rotating frame is
\begin{equation}
f_{1+}^{(B)}(\varepsilon_r) = \frac{\Sigma_c^{(1)}}{2 \pi a \sqrt{\Omega_0^2 -
\Omega^2} \sqrt{2\varepsilon_r}},
\end{equation}
and it can be derived following the same procedure used to find
$f_{1+}^{(A)}(\varepsilon)$ or by the direct application of  eq.
(\ref{metkal2}).

Finally, one can come back to the original frame through the relation between
$\varepsilon_r$, $\varepsilon$ and $L_z$ to obtain
\begin{equation}
f_{1}^{(B)}(\varepsilon,L_{z}) = \frac{\Sigma_c^{(1)} \left[2 (\varepsilon +
\Omega L_z) - \Omega^2 a^2 \right]^{-1/2}}{2 \pi a \sqrt{\Omega_0^2 -
\Omega^2}}, \label{DFm1B1}
\end{equation}
which is totally equivalent to the  \cite{BT} DF for the Kalnajs disk.
Its contours, showed in figure \ref{fig:DFm1C} for different values of $\Omega$, reveals that the probability to
find stars with $\varepsilon < \Omega^{2} a^{2}/2 - \Omega L_{z}$ is zero, while
it has a maximum when $\varepsilon \gtrsim \Omega^{2} a^{2}/2 - \Omega L_{z}$
(this defines the white strip shown in the figure) and decreases as
$\varepsilon$ increases. We take $\Omega=-\pi/4$ in figure \ref{fig:DFm1C}(a), $\Omega=-\pi/16$ in figure \ref{fig:DFm1C}(b), $\Omega=\pi/4$ in figure \ref{fig:DFm1C}(c) and $\Omega = \pi/16$ in figure \ref{fig:DFm1C}(d). Nevertheless, this $f_{1}^{(B)}(\varepsilon,L_{z})$ is
quiet unrealistic as it represent a state with $\langle v_\varphi \rangle =
\Omega R$ and it means that the behavior of the system, in disagreement with the
observations, behaves like a rigid solid.

However, we can generate a better DF if we took only the even part of
(\ref{DFm1B1}) and using the maximum entropy principle to obtain
\begin{equation}
\tilde{f}_{1}^{(B)}(\varepsilon,L_{z}) = \frac{2
f_{1+}^{(B)}(\varepsilon,L_{z})}{1 + e^{-\alpha L_{z}}}.\label{DFm1B}
\end{equation}
As it is shown in figure \ref{fig:DFm1B}, where we plot the contours of $\tilde{f}_{1}^{(B)}$ for
$\Omega=\pi/4$ with $\alpha=1$ in figure \ref{fig:DFm1B}(a) and $\alpha=10$ in figure \ref{fig:DFm1B}(b), and for $\Omega=\pi/16$ with
$\alpha=1$ in figure \ref{fig:DFm1B}(c), and $\alpha=10$ in figure \ref{fig:DFm1B}(d), there is a zone of zero probability,
just in the intersection of black zones produced by $\tilde{f}_{1}^{(B)}(\varepsilon,
\pm L_z)$, and there are also two maximum probability stripes. The variation of
the $\Omega$ parameter leads to the change of inclination of the maximum
probability stripes and it is easy to see that the DF would be invariant under
the sign of $\Omega$, by the definition of the even part. Furthermore, $\alpha$
plays a similar role than in figure \ref{fig:DFm1A}, increasing the probability
of finding stars with high $L_z$ as $\alpha$ increases and vice versa.

\subsection*{DFs for the $m=2$ disk.}\label{sec:dfm2}

Working in a rotating frame we found that the relative potential is given by
\begin{equation}
\begin{split}
\Psi_{2r}(R) = \frac{(a^2 - R^2)}{128 a^5} & \left[45 G M \pi (a^2 - R^2)
\right. \\
&\left. + 30 G M \pi a^2 - 64 a^5 \Omega^2 \right], \label{psi21}
\end{split}
\end{equation}
while the surface mass density is given by (\ref{densidad}) when $m=2$. This
case is a little more complicated than the usual Kalnajs disk, because the
analytical solution of the pseudo-volume density cannot be performed with total
freedom. We will need to operate in a rotating frame conveniently chosen in such
a way that the relation between the mass surface density and the relative
potential becomes simpler.

From eq. (\ref{psi21}) it is possible to see that if we choose the angular
velocity as
\begin{equation}
\Omega = \pm \sqrt{\frac{15 G M \pi}{32a^3}}, \label{omega1}
\end{equation}
the relative potential is reduced to
\begin{equation}
\Psi_{2r}(R) = \frac{45 G M \pi}{128 a^5} \left (a^2 - R^2 \right)^2.
\end{equation}
Now, we can express the surface mass density easily in terms of the
relative potential by the relation
\begin{equation}
\Sigma_2(\Psi_{2r}) = \Sigma_c^{(2)} \left(\frac{128 a}{45 G M \pi} \Psi_{2r}
\right)^{3/4},
\end{equation}
and the integral for the pseudo-volume density can be performed and we obtain
\begin{equation}
\hat{\rho}(\Psi_{2r}) = \frac{\sqrt{2 \pi } \Sigma_c^{(2)}
\Gamma(7/4)}{\Gamma(9/4)} \left(\frac{128 a}{45 G M \pi} \right)^{3/4}
\Psi_{2r}^{5/4}.
\end{equation} \

Using then eq. (\ref{JO}), the resulting even part of the DF in the
rotating frame and the full DF in the original frame are
\begin{equation}
f_{2+}^{(A)}(\varepsilon_r) = \kappa \varepsilon_r^{-1/4},
\end{equation}
and
\begin{equation}
f_{2}^{(A)}(\varepsilon,L_z) = \kappa \left(\varepsilon + \Omega L_z -
\begin{matrix} \frac{1}{2}\end{matrix} \Omega^{2} a^{2} \right)^{-1/4},
\label{DFm2A1}
\end{equation}
respectively, where $\Omega$ could take the values according to (\ref{omega1})
and $\kappa$ is the constant given by
\begin{equation}
\kappa=\frac{3\Sigma^{(2)}_c}{8\pi} \left(\frac{128 a}{45 G M
\pi}\right)^{3/4}.
\end{equation}
Moreover, using the same arguments given in section (\ref{sec:dfm2}), it's
convenient to take the even part of (\ref{DFm2A1}) and obtain a new DF using eq.
(\ref{DFgen1}), which is given by
\begin{equation}
\tilde{f}_{2}^{(A)}(\varepsilon,L_{z}) =
\frac{2f_{2+}^{(A)}(\varepsilon,L_{z})}{1 + e^{-\alpha L_{z}}}. \label{DFm2A}
\end{equation} \

A more general case can be derived without the assumption (\ref{omega1})
if we use the Kalnajs method in order to avoid the pseudo-volume
density integral. Here we work in terms of the spheroidal oblate
coordinates to obtain more easier the relation between the mass
surface density and the relative potential, which can be expressed
as
\begin{equation}
\Psi_{2r}(\eta)=\frac{45 G M \pi  }{128 a}\eta ^4+\frac{15 G M \pi
-32 a^3 \Omega ^2}{64 a}\eta ^2 .
\end{equation}
Now, it is possible to rewrite this expression as
\begin{equation}
\Psi_{2r}(\eta)=(\kappa_1\eta^2+\kappa_2)^2+\kappa_3,
\end{equation}
with
\begin{eqnarray}
&&\kappa_1 = \sqrt{\frac{45 G M \pi}{128 a}}, \\
&&\kappa_2 = \sqrt{\frac{128 a}{45 G M \pi}}\frac{15 G M \pi -32 a^3 \Omega
^2}{128 a}
\end{eqnarray}
and
\begin{equation}
\kappa_3 = - \frac{128 a}{45 G M \pi} \left(\frac{15 G M \pi - 32 a^3 \Omega
^2}{128 a}\right)^2.
\end{equation} \

\begin{figure*}
\centering \epsfig{width=5.6in,file=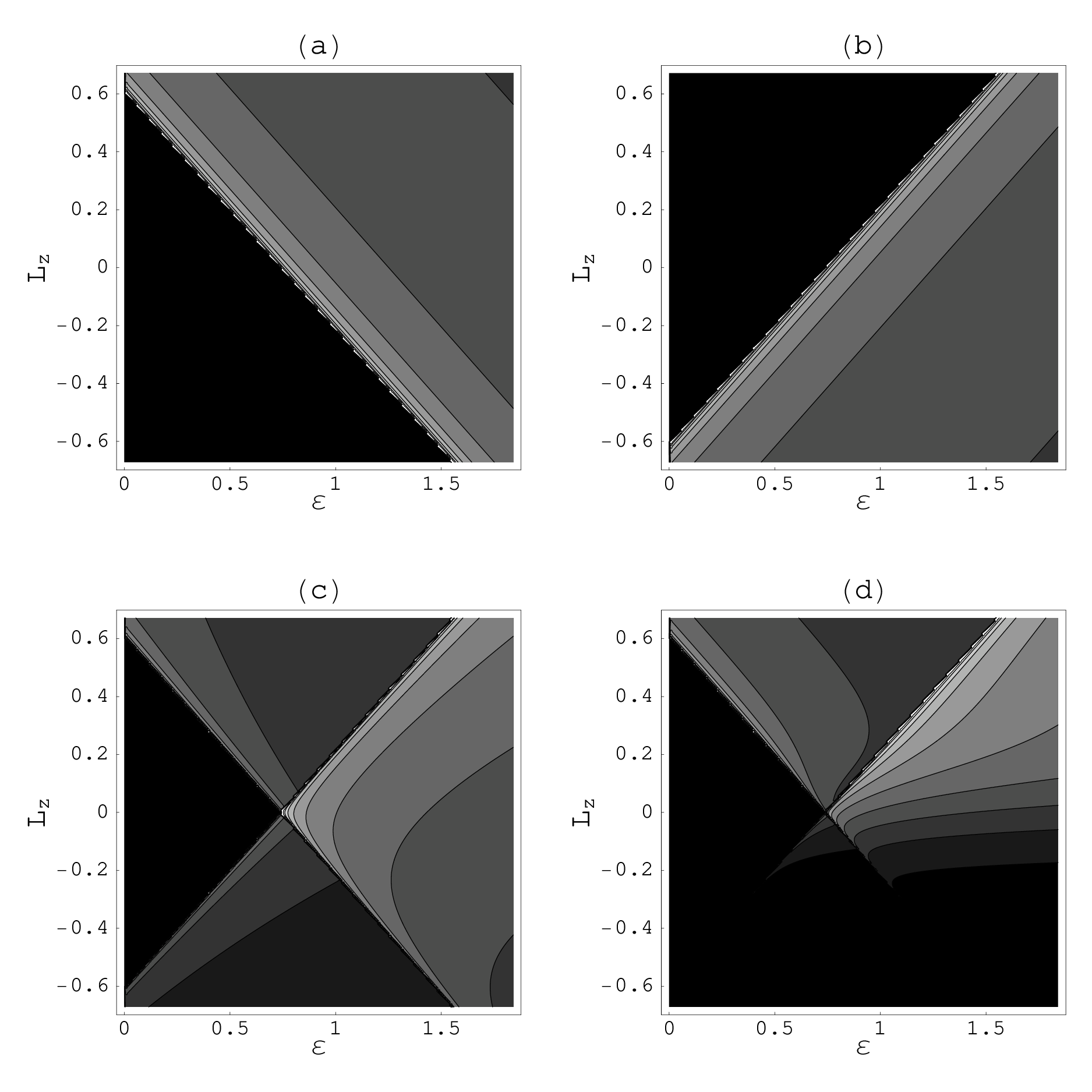} \caption{In (a) and
(b) we show the contours of $f_{2}^{(A)}$, given by (\ref{DFm2A1}),
for $\Omega=\sqrt{15 \pi/32}$ and $\Omega=-\sqrt{15 \pi/32}$,
respectively. The corresponding contours of $\tilde{f}_{2}^{(A)}$, given by
(\ref{DFm2A}), are plotted in (c) for $\alpha=1$, and (d) for $\alpha=10$.
Larger values of the DF corresponds to lighter
zones.}\label{fig:DFm2A}
\end{figure*}

\begin{figure*}
\centering 
\epsfig{width=5.6in,file=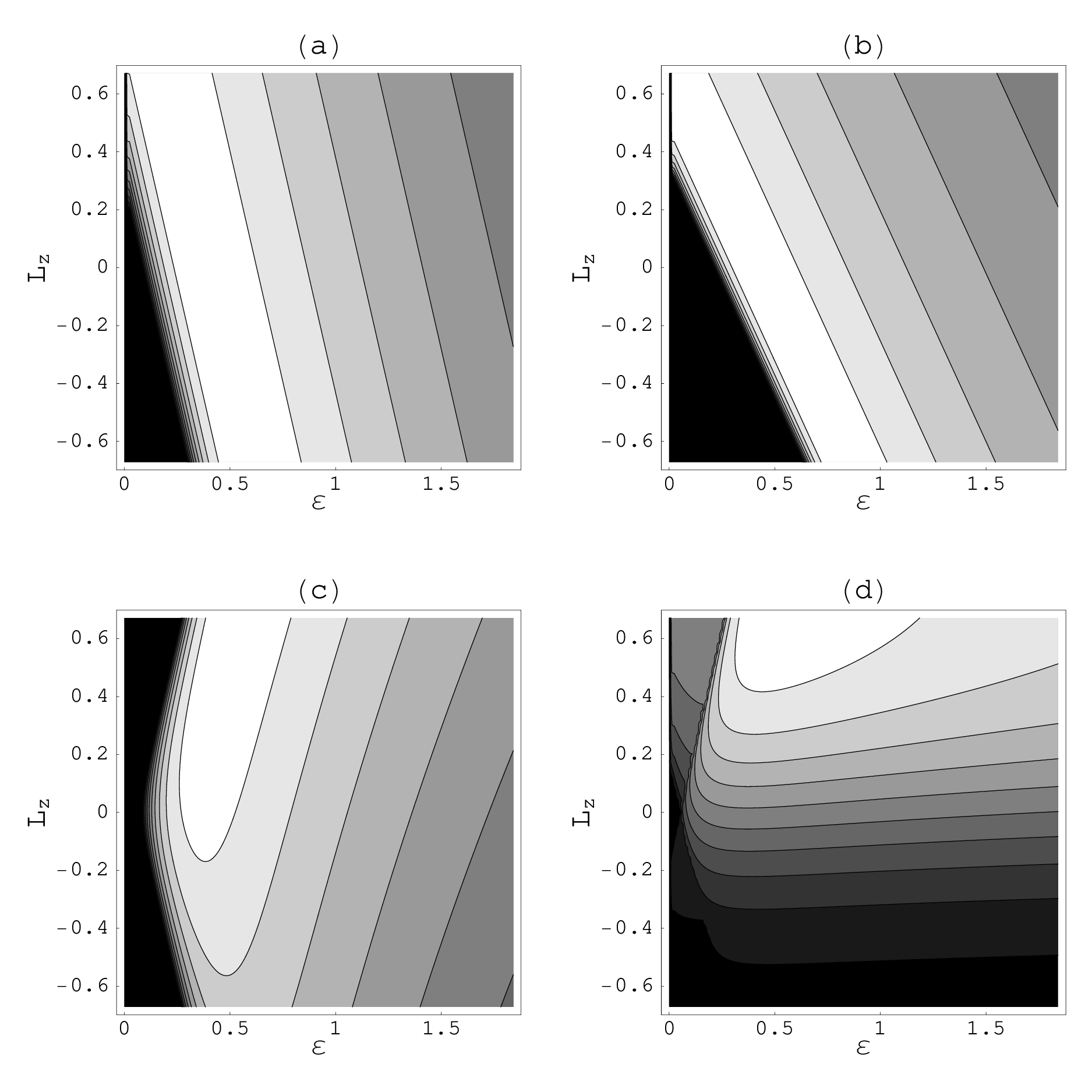}
\caption{In (a) and (b) we show the contours of $f_{2}^{(B)}$, given by
(\ref{DFm2B1}), for $\Omega=\pi/10$ and $\Omega=\pi/5$, respectively. The
remaining figures exhibit the corresponding contours of $\tilde{f}_{2}^{(B)}$, given by
(\ref{DFm2B}), for $\Omega=\pi/10$ with (c) $\alpha=0.1$ and (d) $\alpha=5$. Larger
values of the DF corresponds to lighter zones.}\label{fig:DFm2B}
\end{figure*} 

Then, as $\Sigma$ can be expressed in terms of $\eta$ in the form
\begin{equation}
\Sigma_{m}(\eta) = \Sigma_{c}^{(m)}\eta^{2m-1}, \label{denseta}
\end{equation}
the relation between $\Sigma_{2}$ and $\Psi_{2r}$ is
\begin{equation}
\Sigma_2 = \Sigma_c^{(2)} \left(\frac{\sqrt{\Psi_{2r} - \kappa_3} -
\kappa_2}{\kappa_1} \right)^{3/2}.
\end{equation}
Now, by using eq. (\ref{metkal2}), we obtain
\begin{equation}
f_{2}^{(B)}(\varepsilon_r) = \frac{3 \Sigma_c}{8 \pi k_1^{3/2}}
\left[\frac{\sqrt{\varepsilon_r - k_3} - k_2}{\varepsilon_r - k_3}\right]^{1/2},
\end{equation}
and the result in the original frame is
\begin{equation}
f_{2}^{(B)}(\varepsilon,L_z) = \frac{3 \Sigma_c}{8 \pi k_1^{\frac{3}{2}}} \left[
\frac{(\varepsilon + \Omega L_z - \frac{\Omega^2 a^2}{2} - k_3)^{\frac{1}{2}} - k_2}{\varepsilon
+ \Omega L_z - \frac{\Omega^2 a^2}{2} - k_3} \right]^{\frac{1}{2}}. \label{DFm2B1}
\end{equation}
Obviously, this DF is the same as (\ref{DFm2A1}) when the condition
(\ref{omega1}) is satisfied. Finally, by eq. (\ref{DFgen1}), the resulting DF
with maximum entropy is given by
\begin{equation}
\tilde{f}_{2}^{(B)}(\varepsilon,L_{z}) =
\frac{2f_{2+}^{(B)}(\varepsilon,L_{z})}{1 + e^{- \alpha L_{z}}}. \label{DFm2B}
\end{equation} \

We can see the behavior of these DFs in figures \ref{fig:DFm2A} and
\ref{fig:DFm2B}. In figures \ref{fig:DFm2A}(a) and \ref{fig:DFm2A}(b) we show
the contours of $f_{2}^{(A)}$ for the two rotational states given by eq. 
(\ref{omega1}). Such DF is maximum over a narrow diagonal strep, near to the
zero probability region, and the probability decreases as $\varepsilon$ increases, similarly to the case showed in figure \ref{fig:DFm1B}. The corresponding contours of $\tilde{f}_{2}^{(A)}$, given by eq. 
(\ref{DFm2A}), are plotted in \ref{fig:DFm2A}(c), for $\alpha=1$, and \ref{fig:DFm2A}(d), for $\alpha=10$.

The DFs for stellar systems characterized by different $\Omega$ are shown in figures
\ref{fig:DFm2B}(a) and \ref{fig:DFm2B}(b), where we plot the contours of
$f_{2}^{(B)}$ for the two rotational states given by
(\ref{omega1}). Note that $f_{2}^{(B)}$ equals to $f_{2}^{(A)}$ when $\Omega$ is given by
(\ref{omega1}). In this case the DF varies more rapidly as $\Omega$ decreases,
originating narrower bands. The corresponding contours of
$\tilde{f}_{2}^{(A)}$ and $\tilde{f}_{2}^{(B)}$ for different values of the
parameter $\alpha$ are shown in figures \ref{fig:DFm2B}(c) and \ref{fig:DFm2B}(d), showing a similar behavior than $\tilde{f}_{1}^{(B)}$. We take $\alpha = 1$ for \ref{fig:DFm2B}(c) and 
$\alpha = 10$ for \ref{fig:DFm2B}(d).
 
\subsection*{DFs for the $m=3$ disk.}\label{sec:dfm3}

Once again, if we want to use the Kalnajs method, it is necessary to
derive the relation $\Sigma_3(\Psi_{3r})$ and, according to
(\ref{denseta}), it is posible if we can invert the equation of the
relative potencial, which in this case is given by

\begin{equation}
\begin{split}
\Psi_{3r}(\eta) = &\frac{175 G M \pi}{512 a} \eta^6 + \frac{105 G M \pi}{512 a}
\eta^4 \\ &+ \frac{105 G M \pi - 256 a^3 \Omega^2}{512 a} \eta^2,
\end{split}\label{psi3}
\end{equation}

in order to obtain $\eta(\Psi_{3r})$. To solve it, we must deal with
a cubic equation and with its non-trivial solutions; fortunately, we
still have $\Omega$ as a free parameter.

One can easily note that it
is possible to write (\ref{psi3}) as

\begin{equation}
\Psi_{3r}(\eta) = (\kappa_1 \eta^2 + \kappa_2)^3 + \kappa_3, \label{psi32}
\end{equation}

with
\begin{eqnarray}
&&\kappa_1 = \left(\frac{175 G M \pi  }{512 a}\right)^{1/3}, \\
&&\kappa_2 = \left(\frac{7 \pi G M}{2560 a}\right)^{1/3}, \\
&&\kappa_3 = - \frac{7 G M \pi}{2560 a},
\end{eqnarray}

and $\Omega$ has be chosen as

\begin{equation}
\Omega = \pm \sqrt{\frac{21 G M \pi}{64 a^3}}.
\end{equation} \

\begin{figure*}\centering
\epsfig{width=5.6in,file=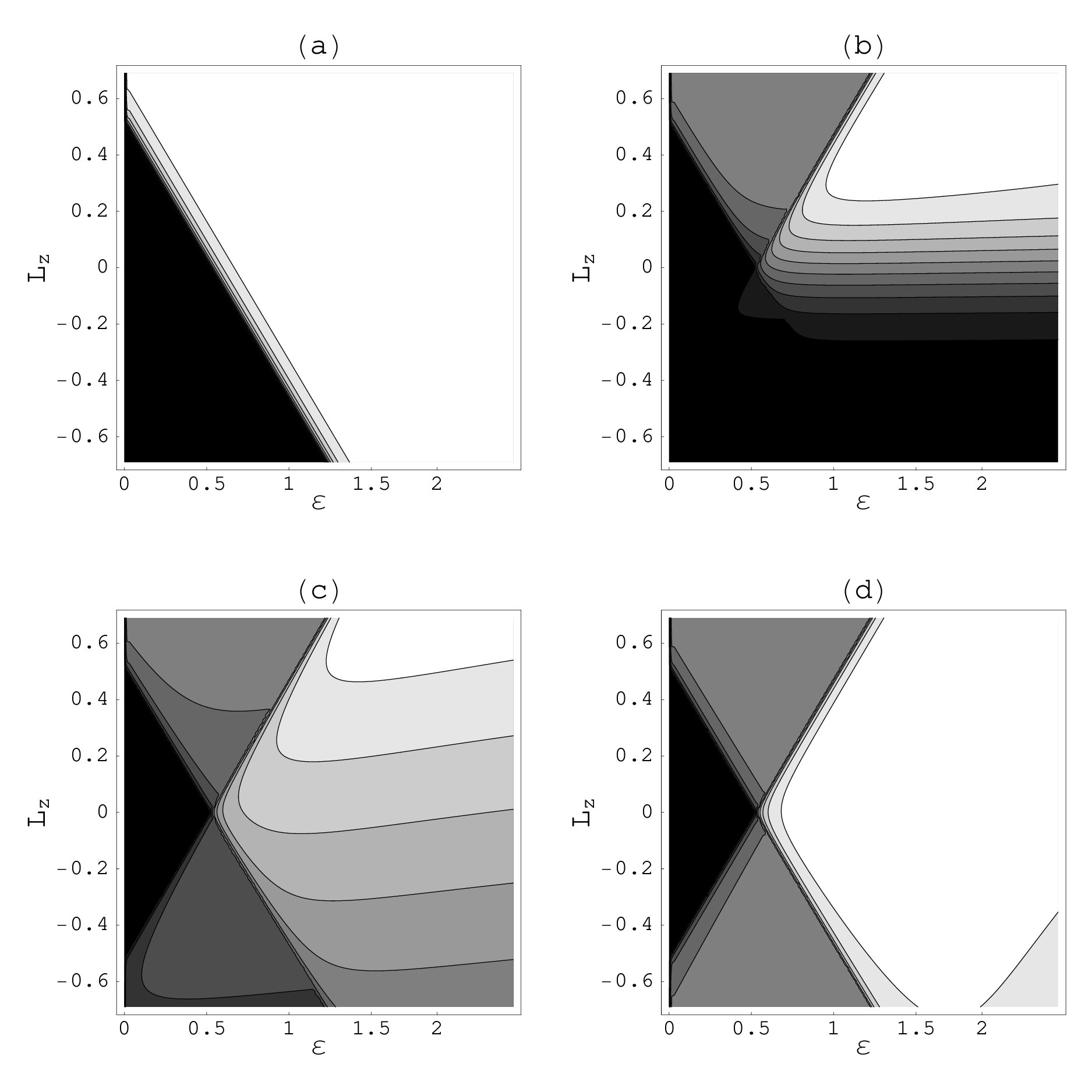}
\caption{In (a) we show the contours of $f_{3}$ given by (\ref{DFm3A1}) and
taking the positive value of $\Omega$. The remaining figures exhibit the
contours of $\tilde{f}_{3}$, given by (\ref{DFm3A}), for (b) $\alpha=10$, (c)
$\alpha=1$, and (d) $\alpha=0.1$. Larger values of the DF corresponds to lighter
zones.}\label{fig:DFm3A}
\end{figure*}

Now, by replacing (\ref{denseta}) into (\ref{psi32}), we obtain

\begin{equation}
\Sigma_3 = \Sigma_c^{(3)} \left(\frac{(\Psi_{3e} - \kappa_3)^{1/3} -
\kappa_2}{\kappa_1}\right)^{5/2},
\end{equation}

and, by using eq. (\ref{metkal2}),
\begin{equation}
f_3(\varepsilon_r) = \frac{5\Sigma_c^{(3)} \left((\varepsilon_r -
\kappa_3)^{1/3} - \kappa_2 \right)^{3/2}}{12 \pi \kappa_1^{5/2}(\varepsilon_r -
\kappa_3)^{2/3}}.
\end{equation}
Coming back to the original frame, the result is
\begin{equation}
f_3(\varepsilon,L_z) = \frac{5 \Sigma_c^{(3)} \left((\varepsilon + \Omega L_z -
\frac{\Omega^2 a^2}{2} - \kappa_3)^{\frac{1}{3}} - \kappa_2 \right)^{\frac{3}{2}}}{12 \pi \kappa_1^{5/2}
\left(\varepsilon + \Omega L_z - \frac{\Omega^2 a^2}{2} - \kappa_3 \right)^{\frac{2}{3}}},
\label{DFm3A1}
\end{equation}
while the respective DF with maximum entropy is
\begin{equation}
\tilde{f}_{3}(\varepsilon,L_{z}) = \frac{2f_{3+}(\varepsilon,L_{z})}{1 + e^{-
\alpha L_{z}}}. \label{DFm3A}
\end{equation}

In figure \ref{fig:DFm3A}(a) we show the contour of  $f_{3}$, while in figures
\ref{fig:DFm3A}(b), \ref{fig:DFm3A}(c) and \ref{fig:DFm3A}(d) are plotted the
contours of $\tilde{f}_{3}$ for different values of $\alpha$. We can see that
the behavior of these DFs is opposite to the previous cases. As the Jacobi's integral icreases, the DF also increases.

\subsection*{DFs for the $m=4$ disk.}\label{sec:dfm4}

As we saw in section \ref{sec:dfm3}, in order to find a DF using the Kalnajs
method, we must find $\eta$ as a function of the relative potential
in order to obtain $\Sigma(\Psi_{r})$. For the $m = 4$ disk, the relative potential can be
expressed as

\begin{equation}
\begin{split}
\Psi_{4r} = &\frac{11025 G M \pi}{32768 a} \eta^8 + \frac{1575 G M \pi }{8192 a}
\eta^6 \\ &+ \frac{2835 G M \pi}{16384 a} \eta^4 + \frac{1575 G M \pi - 4096 a^3
\Omega^2}{8192 a} \eta^2.
\end{split}\label{psir4}
\end{equation} \

Now, although we have to deal with a quartic equation, it is possible to
rewrite (\ref{psir4}) as

\begin{equation}
\Psi_{4r} = [(\kappa_1 \eta^2 + \kappa_2)^2 + \kappa_3 ]^2 + \kappa_4,
\end{equation}

where
\begin{eqnarray}
&&\kappa_{1} = \left(\frac{11025 \pi G M}{32768 a}\right)^{1/4}, \\
&&\kappa_{2} = \left(\frac{225 \pi G M}{1605632 a}\right)^{1/4}, \\
&&\kappa_{3} = \left(\frac{81 \pi G M}{6272 a}\right)^{1/2}, \\
&&\kappa_{4} = - \frac{25281 \pi G M}{1605632 a}, 
\label{cnostDFm4A2}
\end{eqnarray}

and $\Omega$ must to be chosen as

\begin{equation}
\Omega = \pm \sqrt{\frac{135 G M \pi}{448 a^3}}.
\end{equation} \

\begin{figure*}
\centering
\epsfig{width=5.6in,file=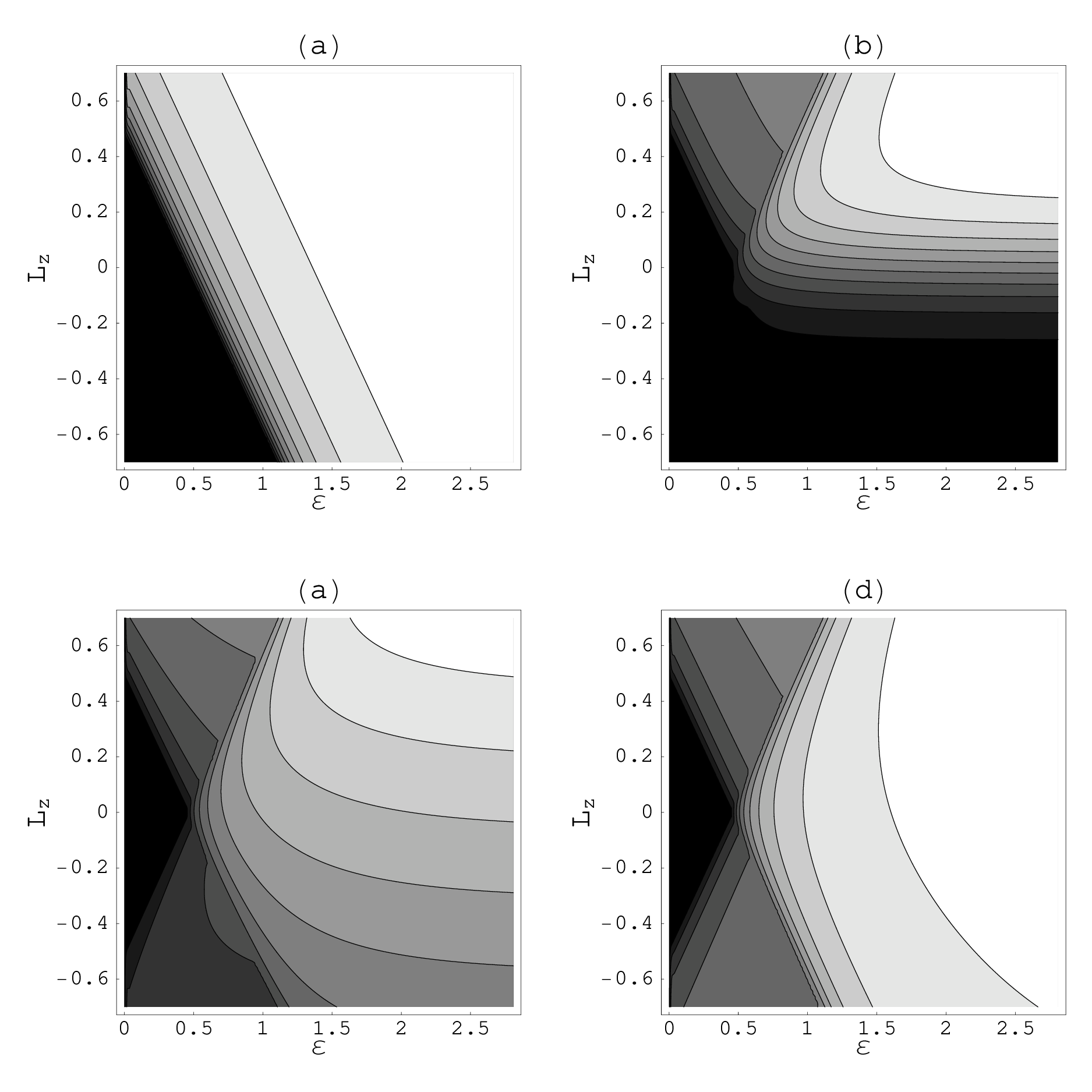}
\caption{In (a) we show the contours of $f_{4}$ given by (\ref{DFm4A1}) and
taking the positive value of $\Omega$. The remaining figures exhibit the
contours of $\tilde{f}_{4}$, given by (\ref{DFm4A}), for (b) $\alpha=10$ (c)
$\alpha=1$ and (d) $\alpha=0.1$. Larger values of the DF corresponds to lighter
zones.}\label{fig:DFm4A}
\end{figure*}

Finally, by using (\ref{denseta}) and (\ref{psi32}), we find the
expression

\begin{equation}
\Sigma_4 = \Sigma_c^{(4)} \left[ \frac{\sqrt{\sqrt{\Psi_{4r} - \kappa_4} - \kappa_3} - \kappa_2}{\kappa_1} \right]^{7/2},
\end{equation}

which, by means of eq. (\ref{metkal2}), can be used to derive the even part of
the DF in the rotating frame,
\begin{equation}
f_4(\varepsilon_r) = \frac{7 \Sigma_c^{(4)} \left[ \sqrt{\sqrt{\varepsilon_r -
\kappa_4} - \kappa_3} - \kappa_2 \right]^{5/2} }{16 \pi \kappa_1^{7/2}
\sqrt{(\sqrt{\varepsilon_r - \kappa_4} - \kappa_3)(\varepsilon_r - \kappa_4)}}.
\end{equation}
Therefore, the corresponding DF in the original frame is given by
\begin{equation}
f_{4}(\varepsilon,L_{z}) = \frac{7 \Sigma_{c}^{(4)}[g(\varepsilon, L_{z}) -
\kappa_{2}]^{5/2}}{16 \pi \kappa_{1}^{7/2} g(\varepsilon,
L_{z})[g^{2}(\varepsilon, L_{z}) + \kappa_{3}]},\label{DFm4A1}
\end{equation}
where

\begin{equation}
g(\varepsilon, L_{z}) = \sqrt{\sqrt{\varepsilon + \Omega L_{z} - \Omega^{2}
a^{2}/2 - \kappa_{4}} - \kappa_{3}}, \label{DFm4A2}
\end{equation}

and the respective DF with maximum entropy is given by
\begin{equation}
\tilde{f}_{4}(\varepsilon,L_{z}) = \frac{2f_{4+}(\varepsilon,L_{z})}{1 + e^{-
\alpha L_{z}}}.\label{DFm4A}
\end{equation}
In figure \ref{fig:DFm4A}(a) we show the contour of  $f_{4}$, while in figures
\ref{fig:DFm4A}(b), \ref{fig:DFm4A}(c) and \ref{fig:DFm4A}(d) are plotted the
contours of $\tilde{f}_{4}$ for different values of $\alpha$. As we can see, the
behavior is analogous to the showed at figure \ref{fig:DFm3A}.

\section*{Concluding Remarks}\label{sec:conc}

We presented the derivation of two-integral equilibrium DFs for some members of
the family of disks previously obtained by \cite{GR}. Such two-integral DFs were 
obtained, esentially, by expresing them as functionals of the Jacobi integral,
as it was sketched in the formalism developed by \cite{KAL2}. Now, since such
formalism demands that the surface mass density can be written as a
potential-dependent function, the above procedure can only be implemented for the
first four members of the family, the disks with $m=1,2,3,4$. Indeed, the procedure 
requires that the expression given the relative potential $\Psi_r$ as a function
of the spheroidal variable $\eta$ can be analytically inverted in order to express
the surface mass density $\Sigma$ as a function of
the relative potential. So,  we can do this in a simple way for the disks with
$m = 1$ to $4$. However, when $m > 4$ we must to solve an equation of grade
larger than four, whose analytical solution do not exists.

For the first two members of the family, the disks with $m=1,2$, we also use the
method introduced by \cite{JO} in order to find the even part of the DF
and then, by introducing the maximum entropy principle, we can determines the
full DF. This procedure was also used for the other three disks, starting from
the \cite{KAL2} method, so defining another class of two-integral DFs. Such kind of
DFs describes stellar systems with a preferred rotational state, characterized
by the parameter $\alpha$. This paper can be considered as a natural complement
of the work previously presented by \cite{GR} and \cite{RLG}, where the PDP
formulation and the kinematics, respectively, of the disks were analyzed. Now,
by the construction of the corresponding two-integral DFs, the first four
members of this family can be considered as a set of self-consistent stellar
models for axially symmetric galaxies.

{\bf Acknowledgments.} GAG was supported in part by VIE-UIS, under grants number 1347 and 1838, and COLCIENCIAS, Colombia, under grant number 8840.

{\bf Conflict of interest.} The authors declare that they have no conflict of interest.

\bibliographystyle{chicago}

\renewcommand{\refname}

\end{multicols}
\end{small}
\end{document}